\documentclass{vgtc}                          

\ifpdf
  \pdfoutput=1\relax                   
  \usepackage{graphicx}                
  \DeclareGraphicsExtensions{.pdf,.png,.jpg,.jpeg} 
\else
  \usepackage{graphicx}                
  \DeclareGraphicsExtensions{.eps}     
\fi%

\graphicspath{{figures/}{pictures/}{images/}{./}} 

\usepackage{microtype}                 
\usepackage{multirow}
\usepackage{subfigure}
\PassOptionsToPackage{warn}{textcomp}  
\usepackage{textcomp}                  
\usepackage{mathptmx}                  
\usepackage{times}                     
\usepackage{comment}
\usepackage{cite}                      
\usepackage{tabu}                      
\usepackage{booktabs}                  
\usepackage{amsmath}
\usepackage{xspace} 
\usepackage{enumitem}
\usepackage{multirow}
\usepackage{verbatim}
\usepackage{amssymb}

\usepackage{colortbl}

\newcommand{\techname}{\textsc{CalTrend}}

\onlineid{1045}

\vgtccategory{Research}

\vgtcinsertpkg

\title{Beyond the Mirror: Personal Analytics through Visual Juxtaposition with Other People's Data}

\author{Sungbok Shin\thanks{e-mail: sungbok.shin@inria.fr. The work was done when Sungbok and Sunghyo were affiliated with Korea University. Sungbok and Sunghyo contributed equally to this research. }\\ %
        \scriptsize Inria%
\and Sunghyo Chung\thanks{shawn.hyo@kakaocorp.com. }\\ %
     \scriptsize Kakao Corp. %
\and Hyeon Jeon\\ %
    \scriptsize SNU %
\and Hyunwook Lee\\ %
    \scriptsize UNIST %
\and Minje Choi\\ %
    \scriptsize Amazon %
\and Taehun Kim\\ %
    \scriptsize Samsung Electronics %
\and Jaehoon Choi\\ %
    \scriptsize Konolabs. Inc. %
\and Sungahn Ko\\ %
    \scriptsize POSTECH
\and Jaegul Choo\thanks{e-mail: jchoo@kaist.ac.kr. Jaegul Choo is the Corresponding Author. }\\ %
    \scriptsize KAIST} %

\abstract{
An individual's data can reveal facets of behavior and identity, but its interpretation is context dependent. 
We can easily identify various self-tracking applications that help people reflect on their lives. 
However, self-tracking confined to one person's data source may fall short in terms of objectiveness, and insights coming from various perspectives. 
To address this, we examine how those interpretations about a person's data can be augmented when the data are juxtaposed with that of others using anonymized online calendar logs from a schedule management app. 
We develop \techname{}, a visual analytics system that compares an individual’s anonymized online schedule logs with using those from other people. 
Using \techname{} as a probe, we conduct a study with two domain experts, one in information technology and one in Korean herbal medicine. 
We report our observations on how comparative views help enrich the characterization of an individual based on the experts' comments.
We find that juxtaposing personal data with others' can potentially lead to diverse interpretations of one dataset shaped by domain-specific mental models.
} 

\begin{document}


\firstsection{Introduction}


\maketitle
\label{sec:intro}
\begin{table*}[!t]
    \caption{\textbf{Works on calendar data analysis.}
    We categorize the 18 papers that utilize schedule logs, or calendar data of a person for visualization, or use calendar-based visualizations. We look at the target task of each work, target users, and whether there was comparison of one's data with others. 
    The papers are listed in chronological order.
    We observe that there is not much work that focuses on providing analysis based on comparing with other users.}
    \label{table:taxonomy_table}
    \centering
    \scalebox{0.97}{%
    \footnotesize
    \begin{tabular}{clcccc} 
    \toprule
     & \textbf{Title (shortened)} & \textbf{Year} & \textbf{Target Task} & \textbf{Target users} & \textbf{Content comparison with others} \\

    \midrule \midrule
    \rowcolor{gray!10} 
    1 & Calendar Data Visualization~\cite{vanwijk99calendarvis} & 1999 & Visual analysis  & N/A & No\\
    2 & Augmented Shared Personal Calendars~\cite{tullio02calendaraugment} & 2002 & Event scheduling & Small group & No  \\
    \rowcolor{gray!10} 
    3 & CMRadar~\cite{modi04cmradar} & 2004 & Calendar management & Personal & No\\
    4 & FishCal~\cite{bederson04datelens} & 2004 & Calendar management & Personal & No\\
    \rowcolor{gray!10} 
    5 & Shared Family Calendars~\cite{plaisant06sharedfamilycalendar} & 2006 & Information sharing & Family & Yes \\
    6 & Personal vs. private calendar interfaces~\cite{tomitsch06calendarint} & 2006 & Understanding calendar use & Personal & No \\
    7 & The Calendar is Crucial~\cite{neustaedter09calendar} & 2006 & Calendar data analysis & N/A  & No\\
    \rowcolor{gray!10} 
   8 & Productivity Tools and Business~\cite{leshed11productbusy} & 2011 & Calendar data analysis & N/A & No  \\
    9 & I'd never get out of this !?\$\%\# office~\cite{dugan12calendar} & 2012 & Calendar data analysis & personal & No \\
    \rowcolor{gray!10} 
    10 & Survey on Personal Vis~\cite{huang15personalvisualanalytics} & 2015 & Survey & N/A & N/A \\
    11 & TimeAware~\cite{kim16timeaware} & 2016 & Tracker & personal & No \\
    \rowcolor{gray!10} 
    12 & Calendar.help~\cite{cranshaw17calendarhelp} & 2017 & Event scheduling & Small group& Yes \\ 
    13 & T.Cal~\cite{fu18tcal} & 2018 & Work pattern analysis & Small group & Yes \\
    \rowcolor{gray!10} 
    14 & Pervasive Wall Calendar~\cite{voit18pervasivecal} & 2018 & Calendar augmentation & Personal & No \\
    15 & Neural event scheduling assistant~\cite{kim18kono} & 2018 & Event scheduling & Small group & No \\
    \rowcolor{gray!10} 
    16 & DayClo~\cite{lee20dayclo} & 2020 & Calendar augmentation & Personal & No \\
    17 & Personalized Calendar at a care center~\cite{rodil20spendingtime} & 2020 & Collaborative event scheduling & Small group & No \\
    18 & Shift scheduling system for healthcare workers~\cite{uhde21selfschedulehsw} & 2021 & Shift scheduling & Small group & No \\
    
    \bottomrule
 \end{tabular}
}
\end{table*}


Understanding self stands as a cornerstone for personal and professional development.
Marking the everyday progress, observing changes and evaluating self-improvements enable us to self-reflect and better strategize our ways of living~\cite{lupton2016quantified}.  
Consequently, personal-tracking services for productivity, time management, and health have gained wide popularity.

However, the inherently societal aspects of human life\footnote{\textit{``A man is by nature, a social animal''}, Aristotle, \textit{Politics, Book 1}, section 1253a}\cite{nippert2008home} might imply that considering only an individual's personal data could constrain the interpretative power of personal analytics. 
In many scenarios, comparative context is essential; for instance, while a 20\% salary increase may initially seem substantial, understanding whether it is truly competitive requires knowledge of peers' earnings. 
Similarly, establishing whether consistently sleeping 10 hours per night aligns with typical social norms demands data from the wider population.

Despite the broader potential of personal analytics, existing work remains predominantly centered on analyzing an individual's own historical data. 
This observation is supported by a review of 18 studies on schedule logs and calendar visualization (see \autoref{table:taxonomy_table}), most of which emphasize self-comparison rather than comparative analysis involving other individuals.

In this work, we explore how juxtaposing multiple users’ records can enrich the interpretation of an individual’s own tracked data. 
To that end, we introduce \techname{}, a visual analytics system that employs t-SNE, a dimension reduction technique and topic modeling to provide temporal and textual pattern analysis for both single and multiple online calendar users. 
We then present \techname{} to two professional analysts from different domains (Korean herbal medicine, and human resources management team at an IT company) as a probe, asking them to interpret individual user behavior within the dataset and provide actionable insights.
We utilize anonymized online schedule logs from an online scheduling service.
We report on the findings from the observations that these experts have made while analyzing the data using \techname{}.


\section{Design Requirements for \techname{}}
\label{sec:design_rationales}

\techname{} serves as an analytical probe that illustrates how examining an individual’s data in the context of others’ datasets can yield deeper and more nuanced insights.
For this purpose, we denote four design requirements for \techname{}.

\begin{itemize}[itemsep=0.001cm, topsep=0.1cm, leftmargin=0.9cm]
    \item [\textbf{DR1}] \textbf{Contrastive. } As we aim to compare and contrast the difference between selected groups of the analyst's interest, the tool should be able to effectively highlight the differences.
    \item [\textbf{DR2}] \textbf{Multi-faceted data analysis. } As an analytical tool, the tool should provide analysis of the datasets from multiple angles. 
    \item [\textbf{DR3}] \textbf{Catching different temporal patterns. } In an individual’s schedule logs, various temporal patterns emerge at multiple levels of granularity (for example, weekly, daily, etc.). 
    Hence, for a close investigation, we need to effectively present these temporal patterns. 
    \item [\textbf{DR4}] \textbf{Effective summary of contexts. } Schedule logs often contain extensive textual content. For effective analysis, we need to effectively summarize these texts. 
\end{itemize}

\section{\techname{} Implementation and Design}
\label{sec:calivian_method}

We present the implementation and design of~\techname{}.
First, we present the overview, how the data is processed, functionalities in 

\subsection{\techname{} Overview}

\begin{figure}
\centering\includegraphics[width=0.47\textwidth]{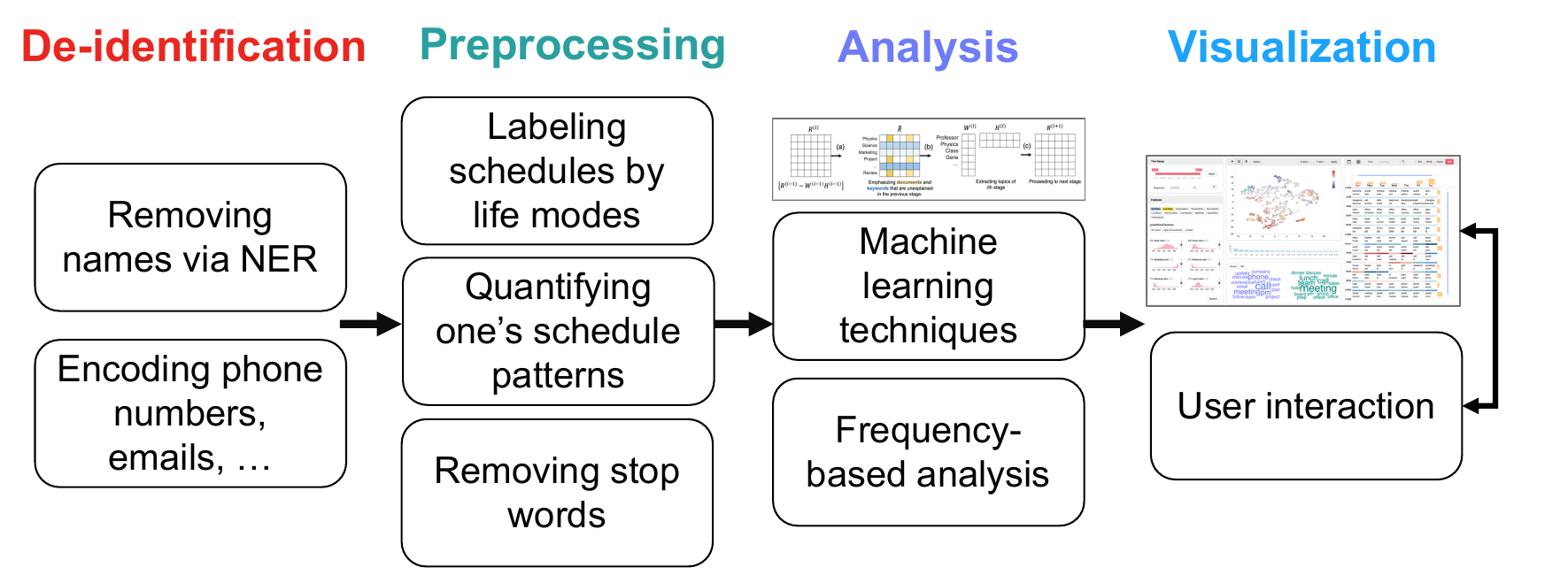}
\caption{\textbf{The overall workflow of \techname{}.} \techname{} is developed in three stages: first, we deidentify information from the schedule logs. Second, we preprocess the data by annotating life modes and defining features reflecting each user's scheduling patterns. Finally, we apply various machine learning (ML) techniques to generate insights, which are integrated into the \techname{} visual analytics system. }
\label{fig:overview}
\end{figure}

\autoref{fig:overview} illustrates the overall workflow of \techname{}. 
We begin by deidentifying the schedule logs, replacing personally identifiable information, such as phone numbers, emails, and addresses—with arbitrary values and removing names. 
Next, we preprocess the data to facilitate the analysis of online calendar users’ life patterns, including labeling schedules by life modes and quantifying relevant features. 
We then apply Machine Learning techniques and frequency-based analyses to extract insights from the data. 
Finally, \techname{} visualizes these results and provides interactive tools to aid in exploring individual user behaviors.

Below, we first provide a brief introduction of online schedule logs (\S~\ref{subsec:sched-logs-intro}), describe the preprocessing steps used (\S~\ref{subsec:preprocessing}), and describe the \techname{} interface (\S~\ref{subsec:interface-explain}, \S~\ref{subsec:detailed_view}). 

\begin{figure*}
\centering\includegraphics[width=0.99\textwidth]{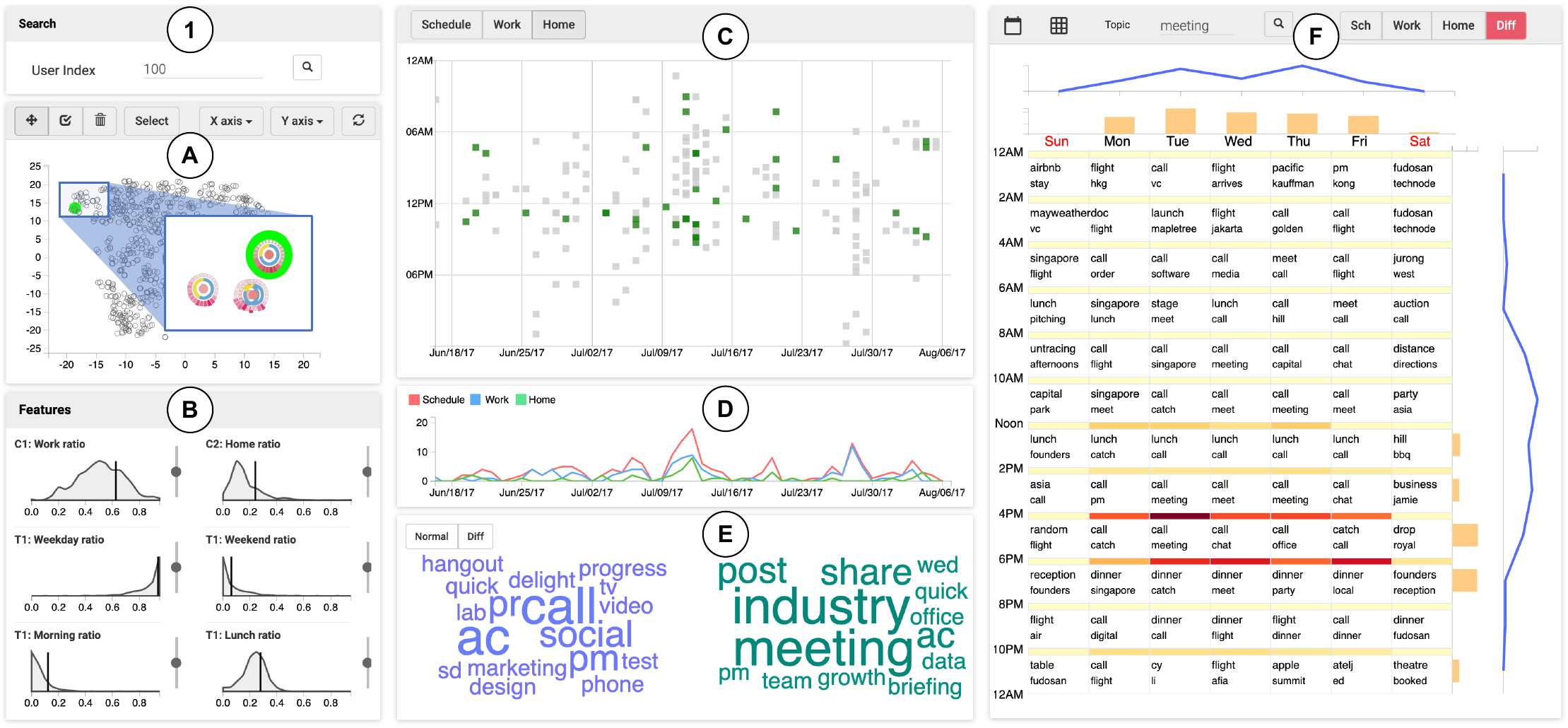}
\caption{\textbf{The \techname{} interface.} \techname{} is a visual analytics system designed to reveal temporal and contextual trends in online calendar logs. 
(A) A scatterplot presents a two-dimensional projection of calendar users, where zooming transforms dots into glyphs encoding schedule frequency, life modes, and total number of events. 
Hovering over a dot displays its associated textual data. 
From (A) and (1), the user can select a user group(or individual) of their interest (see \S~\ref{subsec:interface-explain}). 
(B) The analyst can adjust feature weights to modify the t-SNE projection, with black bars in the histogram indicating the user’s values for each quantified feature. 
(C) An hour-by-day heatmap captures daily scheduling patterns, while (D) a line graph visualizes the cumulative frequency of schedules. 
(E) Two topic-model-based wordclouds highlight key textual themes, and (F) a novel hour-by-week heatmap illustrates repetitive scheduling patterns, offering interactions such as keyword distribution analysis (see \S~\ref{subsec:detailed_view} for more details). 
Overall, these integrated views provide a comprehensive perspective on user behaviors, supporting rich domain-specific insights.}
\label{fig:system-screenshot}
\end{figure*}

\subsection{Schedule Logs from Online Calendar Users}
\label{subsec:sched-logs-intro}

Online schedule logs were provided by Konolabs Inc., an AI-based schedule management company. 
The dataset comprises 1,652,071 schedule entries from 1,025 users in diverse domains (e.g., businesses, startups, education), all of whom primarily use English. 
Personally identifiable information (e.g., names, emails) has been completely deidentified.

Each schedule log consists of three parts: (1) textual data (e.g., summary), (2) temporal data (e.g., start time, time zone), and (3) attendee information. 
The textual data describes the event’s purpose (e.g., meeting or reminder). 
Temporal data details when the event occurred, how long it lasted, and when it was updated. 
Attendee information specifies who created the event and who participated, supplemented by time zone data to ensure accurate scheduling details.

\subsection{Preprocessing Schedule Logs}
\label{subsec:preprocessing}

Before developing the system, we conduct preprocessing steps on online schedule logs, such as (1) labeling schedules into different facets of life and (2) setting features to represent one's online calendar user.

\smallskip\noindent\textbf{Labeling schedules by life facets.} Drawing on the notion of \textit{life facets} or \textit{life modes}, specifically \textit{work} and \textit{home}\cite{turner78differentiation, farnham11facetedlives, nippert2008home, ozenc11lifemodes}, we follow Nippert-Eng’s\cite{nippert2008home} examination of how individuals shape the boundaries between these two domains (DR2). 
Motivated by her findings, we adopt `work' and `home' as our labels.

We collected all schedule summaries and removed stop words, yielding 199,655 bag-of-word keywords. 
Using the Stanford Named Entity Recognizer (NER)\footnote{\url{https://nlp.stanford.edu/software/CRF-NER.html}}, we classified entities, identifying 156,654 as names. 
From the remaining 40,162 keywords, we constructed \textit{concepts}~\cite{park18conceptvector}, sets of semantically related terms, for the two labels. 
Three authors of the paper manually clustered these keywords into \textit{work} and \textit{home}, focusing on those appearing in more than 50 logs and retaining only mutually agreed-upon terms. 
To further validate this set, the Konolabs research team reviewed and removed any ambiguous items. 
As a result, we identified 1,713 work-related keywords that are primarily office-, technology-, or finance-oriented, and 472 home-related keywords, comprising social and private life themes.

Finally, we assigned each schedule a life-mode label (work, home) if its summary contains keywords from the corresponding concept; schedules containing keywords from multiple life modes receive multiple labels. 
Of the labeled schedules, 11.3\% contained more than one label. 
Overall, 63.2\% of schedules were labeled under one or both life modes, with 70.2\% categorized as \textit{work}, 24.3\% as \textit{home}, and 16.7\% remaining unlabeled. 
Additionally, names appeared in nearly 62\% of all schedules.


\smallskip\noindent\textbf{Defining features to represent online calendar users. }
To contextualize each user's temporal patterns, schedule distributions and textual context, we defined 11 representative features (DR2). 
Two features quantify \textit{schedule modifications} and \textit{monthly schedule volume}. 
Another two gauge the ratio of \textit{weekend} to \textit{weekday} events. 
Five additional features track hourly distributions (\textit{morning}, \textit{lunch}, \textit{afternoon}, \textit{evening}, \textit{night}). 
Finally, two features measure the rate of \textit{work} and \textit{home} content within a user's schedule logs. 
They reflect each user’s textual focus.

\subsection{Selection Interfaces}
\label{subsec:interface-explain}

We present two interfaces for selecting online calendar users of the analyst's interest. 
First, the user can search directly for the user of their interests based on the index number (see the `search' in \autoref{fig:system-screenshot}\textcircled{1}). 
Second, analyst can directly choose clusters or groups by selecting members from the map (\autoref{fig:system-screenshot}(A)). 
Once analyst selects the users, they can view the detailed information on them. 
Then, the hour-by-week visualization and the wordcloud visualization, which we explain in \S~\ref{subsec:detailed_view}.
Below we explain in detail the functionalities of the 2-dimensional projection map, where users can select users from various perspectives. 

\smallskip\noindent\textbf{2-dimensional projection map.} We present a two-dimensional projection of online calendar users (\autoref{fig:system-screenshot}(A)), enabling analysts to detect clusters, identify anomalous users, and select areas of interest for detailed inspection. 
Each user is represented by 11 predefined features, which we reduce to two dimensions via t-SNE (DR2). 
By adjusting feature weights or focusing on specific categories (e.g., temporal attributes), analysts can generate varied projection layouts and uncover distinct clusters. 
\autoref{fig:system-screenshot}(B) illustrates the interface for controlling feature weights.
Additionally, users can construct scatterplots by choosing any two of the 11 features as axes. 
For instance, to examine the correlation between work and social ratios, the analyst selects ``work ratio'' for the x-axis and ``social ratio'' for the y-axis. 
Clicking the ``show'' button then displays how user data points are distributed with respect to these two features.

\smallskip\noindent\textbf{Glyphs for individual analysis.} When zoomed in by approximately five mouse-scrolls, the 2D-projection dots transform into glyphs (\autoref{fig:system-screenshot}(A)), offering a perceptual summary of each user’s temporal and life-mode patterns (DR2, DR4). 
The inner circle’s color denotes the total number of schedules, while the ring-shaped middle layer encodes the distribution of life modes (work or home). 
The outer layer is a fan-shaped bar chart divided into one-hour segments, with each segment’s width and color reflecting the frequency of schedules during that hour.

\subsection{Detailed Views on Online Calendar Users  }
\label{subsec:detailed_view}

We present various temporal patterns of schedule logs via (1) hour-by-day heatmap, (2) a hour-by-week heatmap, as well as effective summary of texts using (3) two wordclouds (DR3, DR4). 

\smallskip\noindent\textbf{Hour-by-day heatmap.} The hour-by-day heatmap displays schedules on a two-dimensional plane, with the horizontal axis indicating the date and the vertical axis indicating the time. 
Each schedule is initially depicted as an opaque gray rectangle. 
By selecting the ``work'' or ``home'' button, the analyst can highlight the corresponding schedules in blue or green, respectively. 
Scrolling up or down adjusts the zoom level for more detailed examination.

\smallskip\noindent\textbf{Hour-by-week heatmap visualization.} The hour-weekly heatmap (\autoref{fig:system-screenshot}(F)) effectively reveals recurring weekly and hourly patterns that may not be discernible in \autoref{fig:system-screenshot}(C) (DR3). 
Each column represents a weekday, from Sunday to Saturday, and is divided into 12 two-hour segments. 
Each heatmap cell comprises a top bar indicating schedule frequency (via opacity) and a set of keywords reflecting the most frequent terms for that time slot. 
Analysts can view three variants of the heatmap—(1) all schedules, (2) work-labeled schedules, and (3) home-labeled schedules—and toggle a ``diff'' mode using the red ``diff'' button to highlight frequency differences (DR1). 
Hovering over a cell displays a tooltip with its top 10 keywords, and the number of keywords shown scales with schedule frequency. 
Horizontal and vertical bar charts adjacent to the heatmap illustrate schedule frequency by weekday, while adding keywords above the blue line graph enables exploration of distribution patterns across both weekdays and hours.

\smallskip\noindent\textbf{Two wordclouds.} To generate topical summaries of online calendar user(s), we provide four topic models in our wordcloud view (\autoref{fig:system-screenshot}(F)). 
The first two wordclouds display the top 10 keywords from standard topic modeling of work (blue) and home (green) textual logs, with keyword size indicating relative dominance within each topic. 
Activating the ``diff'' button highlights less frequent yet more distinctive work- or home-labeled terms (DR1). 
When viewing a single user, the wordcloud reflects that individual’s topic models; for multiple users, it aggregates the most frequent keywords from all selected users’ topic models.

\subsection{System Implementation}

The frontend web interface is built using Javascript with jQuery. We utilized the D3 library (V3)~\footnote{https://d3js.org/} to create various visual components and d3-cloud to design word clouds. The backend server and functions are constructed using Python 3.6 with WebSocket protocols~\footnote{https://developer.mozilla.org/en-US/docs/Web/API/WebSocket}.

\section{Study Design}

Here we discuss how the process the two experts took during the study using  \techname{}. 
The first participant is a marketing expert who is a general director in the marketing branch at LG Electronics, and the second is a korean herbal medicine doctor. 
We obtained agreements from these two domain experts about not disclosing the information they obtained from schedule logs.
We asked them to use the tool, look at the distribution, identify the anomalous users, and think of additional methods to characterize online calendar users. 

Each study lasted around one hour and ten minutes. To each we explained the system for 15 minutes, and had the participant test our system for 10 to 15 minutes. 
Then, we discussed the possibilities and limitations of using schedule logs for their own domains. 
Before the experts began using our system, we first asked them to explain in detail how our task of (1) selecting online calendar users and (2) providing temporal, elaborate details about them would be applied to their specific domains. They provided comments in the middle of using the tools. 
From the two experts, we observe that characterizing using others' information can enrich the explanation about oneself, when added with the experts' domain knowledge. 
Second, these experts' identification of special groups/anomalies of interest, can be used to enrich the characterization of these online calendar users. 

\section{Takeaways}

We describe the two main takeaways based on the activities and think-aloud comments from the experts. 
We find that personal information, when juxtaposed with others' data, can be interpreted in diverse ways, and that the mental models used to make sense of such information vary across domains.

\subsection{Domain-specific interpretations of schedules}
Both experts grounded the users' calendar schedules in problems relevant to their respective domains and proposed methods to generate meanings from the data accordingly. 
The marketing expert noted that individuals exhibit distinct lifestyles encompassing culinary preferences, workout routines, and hobbies in calendars. 
He mentioned that leveraging these insights from \techname{} can enable more tailored marketing strategies. 
The medical expert treated the schedules as indicators of daily effort and stress, suggesting their potential application in predicting mental health issues such as depression or burnout. 
He further recommended incorporating additional health-related events, like sleep schedules, to heighten the data's clinical relevance. 

Overall, these diverse perspectives underscore the potential of calendar-based analytics for providing meaningful, domain-specific insights across multiple fields.

\subsection{Identification of Groups of Interests based on domain-specific interpretations}

Domain-specific expert insights guided the identification of specialized user groups within the online calendar data, yielding additional characteristic dimensions. 
Experts explored these groups from the 2D map by adjusting parameters in the cluster view: the marketing expert sought segments distinguished by lifestyle and interests, while the herbal medicine doctor manipulated feature weights to detect anomalous schedules, such as those occurring late at night or at exceptionally high volumes, or those that lie in outliers of charts. 
The marketing expert tried to identify different kinds of life patterns by frequently changing the 2D map views.  
These observations highlight the practical utility of multi-dimensional analytics for domain-specific characterizations.

Depending on the domain, usage of the tool varied significantly. 
After identifying anomalous patterns, the herbal medicine doctor examined potential causes using textual and detailed views, suggesting that such insights could reveal information patients may not openly share. 
Moreover, displaying these distributions could help patients determine whether their behaviors are normative or deviate from the norm. 
By contrast, the marketing expert leveraged these findings to brainstorm product ideas for distinct user segments, aligning this process with the STP (Segmenting, Targeting, and Positioning) framework. 
Specifically, segmenting groups users with shared characteristics, targeting directs resources toward a chosen segment, and positioning involves communicating a product’s value to that group. 
In this manner, \techname{} helped marketing analysts in identifying target groups suited to specific lifestyles. 

These findings reveal that the mental models through which users interpret juxtaposed data vary by domain. 
Understanding these domain-specific mental models is essential for developing tailored visualizations that support self-understanding from multiple professional perspectives. 
We see this as a promising direction for future work. 

\section{Limitations}
Concerns regarding privacy issues were raised by viewing the schedules of others. 
While our setting ensures anonymity, in a small group setting the calendar events may still be sufficient to identify certain users. 
This issue was raised by the marketing expert while discussing the applicability of \techname{} in office settings. 
While this issue is not a concern when limited to work-related calendar events which are often visible between organization members, the incorporation of personal events may cause privacy issues.

Furthermore, we used only one dataset and two domain experts to explore the potential of juxtaposing other people's data to enrich one's personal information. 
However, there exist numerous datasets pertaining to personal data and experts from various domains who can provide varying insights about one's personal information. 
Hence, while our goal was in exploring the possibility of enriching one's understanding by juxtaposing others' information, we think that the next goal for this is to use multiple datasets and various domain experts to further generalize the results.

\section{Conclusion}
\label{sec:conclusion}

In this paper, we aim to understand how juxtaposing other people's data can enrich a person's data. 
To that end, we proposed \techname{}, a visual analytics system composed various visualization components to illustrate the comparative differences between a group and the other on temporal and contextual information, based on the analysis of online schedule logs. 
Finally, we provided an expert interview with 2 domain experts (one in Korean Herbal Medicine, and one in Marketing) to identify how our approach can benefit and enrich personal analytics.

\bibliographystyle{abbrv-doi}

\bibliography{sungbokshin}
\end{document}